\documentclass[aps,prx,reprint,superscriptaddress,showpacs]{revtex4-1}
\usepackage{graphicx}
\UseRawInputEncoding 
\usepackage{amssymb}
\usepackage{color}
\usepackage{amsmath}
\usepackage{ulem}
\usepackage{bm}
\usepackage{mathrsfs}
\newcommand{\be}{\begin{eqnarray}}
\newcommand{\ee}{\end{eqnarray}}
\newcommand{\la}{\langle}
\newcommand{\ra}{\rangle}

\begin{document}

\title{Phase diagram of the extended chequerboard $J-Q$ model}
\author{Jiayou Yin}
\affiliation{School of Physics, Beijing Institute of Technology, Beijing 100081, China}
\author{Lu Liu}
\email{liulu96@bit.edu.cn}
\affiliation{School of Physics, Beijing Institute of Technology, Beijing 100081, China}

\begin{abstract}
The chequerboard $J-Q$ model was proposed to describe the direct phase transition from the antiferromagnetic (AFM) state to the plaquette-singlet (PS) solid state observed in SrCu$_2({\rm BO}_3)_2$. In this paper, we present a Monte Carlo study of the ground state of an extended version of this model. For all parameters investigated, we find only a direct first-order phase transitions from the AFM to the PS phase, with no intermediate phase between them. On the transition line, the system exhibits an emergent $O(4)$ symmetry. Furthermore, we find that the Binder ratio of the columnar valence-bond solid state can be used to locate the phase transition. It exhibits a monotonic finite-size scaling behavior, allowing for a precise determination of the transition point.

\end{abstract}

\date{\today}

\maketitle

\section{Introduction}
The investigation of strongly correlated quantum spin systems represents a central endeavor in modern condensed matter physics, driven by the pursuit of exotic states of matter and unconventional critical phenomena~\cite{sachdev2008,kaul2013}. Frustrated magnetic lattices, in particular, serve as ideal platforms for realizing such physics. They host a rich variety of quantum phases, including novel orders, quantum spin liquid (QSL), and unconventional quantum critical point (QCP). A prominent example of the latter is the deconfined QCP (DQCP), which is characterized by fractionalized excitations and emergent gauge fields~\cite{senthil2004-1,senthil2004-2}.

The two-dimensional Shastry-Sutherland (SS) model is one of the most fascinating quantum models~\cite{ssmodel}. It hosts exotic nonmagnetic ground states~\cite{jylee2019,zayed2017,hchen2025,yyang2024,jyang2022,lwang2022,nakano2018,viteritti2025,corboz2025,corboz2013,kliu2024,zdeng2025,akoga2000,jwang2023,maity2024,wyliu2024,jyang2024,xqian2024,nxi2023,akeles2022,lchen2025,cboos2019}, possibly (proximate to) DQCP~\cite{jylee2019,zayed2017,hchen2025,wyliu2024} and QSL~\cite{jyang2022,lwang2022,nakano2018,viteritti2025,corboz2025,kliu2024,maity2024,akeles2022}, as well as exciting possibilities to realize altermagnetism~\cite{ferrari2024,hchen2025} . The SS model is faithfully realized by the quantum magnetic material SrCu$_2({\rm BO}_3)_2$ whose pressure-field-temperature phase diagram is under intensive investigation~\cite{jguo2020,jguo2025,ycui2023,ycui2025,kageyama1999,jimenez2021,mcclarty2017,fogh2024,nomura2023}. 
An interesting property of both the SS model and SrCu$_2({\rm BO}_3)_2$ is the transition from the antiferromagnetic (AFM) state to the plaquette-singlet (PS) solid state. There is no consensus on the nature of this transition. Competing findings include a direct continuous transition~\cite{jylee2019,zayed2017,akoga2000,wyliu2024,jyang2024}, a first-order transition~\cite{corboz2013,zdeng2025,xqian2024,nxi2023,lchen2025,jguo2020,jguo2025,ycui2023,ycui2025}, and  an intervening narrow quantum spin liquid phase between them~\cite{jyang2022,lwang2022,nakano2018,viteritti2025,corboz2025,maity2024,akeles2022}. Notably, the direct continue or first-order transition is thought to be governed by the deconfined quantum criticality~\cite{jylee2019,zayed2017,hchen2025,wyliu2024,ycui2023,ycui2025}.

Due to the sign problem in the SS model, the chequerboard $J-Q$ (CBJQ) model was introduced as a designer Hamiltonian~\cite{Bzhao2019} that is amenable to large-scale quantum Monte Carlo (QMC) simulations in order to imitate the SS model. The CBJQ model hosts a first-order phase transition from AFM to PS with emergent $O(4)$ symmetry. 
On grounds of symmetry and universality, it can be argued that the two models, along with SrCu$_2({\rm BO}_3)_2$, share the same universal physics underlying the AFM-PS transition.
 Thus, the CBJQ model provides a consistent theoretical framework for understanding SrCu$_2({\rm BO}_3)_2$.
 The first-order AFM-PS thermodynamic transition in SrCu$_2({\rm BO}_3)_2$ is predicted by CBJQ model~\cite{Bzhao2019, jguo2020}. The finite-temperature bi-critical behavior predicted by CBJQ model is also verified experimentally~\cite{jguo2025}. In addition, the field-induced AFM-PS transition in SrCu$_2({\rm BO}_3)_2$ is successfully accounted for by the same model~\cite{ycui2023}.

Despite its successes, the CBJQ model has certain limitations. Firstly, the CBJQ model exhibits a strongly discontinuous
AFM-PS transition, whereas 
the SS model suggests a continuous or weakly first-order transition, potentially with an intervening narrow quantum spin liquid phase. 
Secondly, the model stabilizes only the full PS (FPS) state 
but not the empty PS (EPS) state. This absence of FPS-EPS competition contrasts with the strong EPS-FPS competition reported in both the SS model and SrCu$_2({\rm BO}_3)_2$\cite{ycui2025,lchen2025,nxi2023,cboos2019}.

In order to overcome these limitations, we generalize the CBJQ model by introducing independent four-spin coupling strengths on the two types of plaquettes. This generalization extends the parameter space to tune the competition between different plaquette ordering patterns and may bridge the gap between the CBJQ model and the SS physics.
It is a common and powerful methodology in the study of quantum magnetism to explore generalized or extended model, as the extended model can reveal richer phase diagrams, stabilize new quantum states, and provide deeper insights into unresolved critical behavior. For example, various generalizations of the SS model have not only unveiled a wealth of novel physical phenomena but also provided crucial insights into the physics of the original model itself~\cite{maity2024,wyliu2024,xqian2024,nxi2023,akeles2022,lchen2025,cboos2019}. In this paper, we study the ground-state properties of the extended CBJQ model using large-scale stochastic series expansion (SSE) QMC simulations~\cite{sandvik}. 

This paper is organized as follows. In Sec.~\ref{modelandmethod}, we introduce the extended CBJQ model, along with observables calculated. In Sec.~\ref{results}, we present the results: the phase diagram (Sec.~\ref{results}~A), the emergent symmetry at the transition points (Sec.~\ref{results}~B), and the effective order parameter (Sec.~\ref{results}~C). Section \ref{summary} summarizes
the key findings of this study.


\section{Model and observables}
\label{modelandmethod}
\begin{figure}[t]
  \includegraphics[width=45mm,clip]{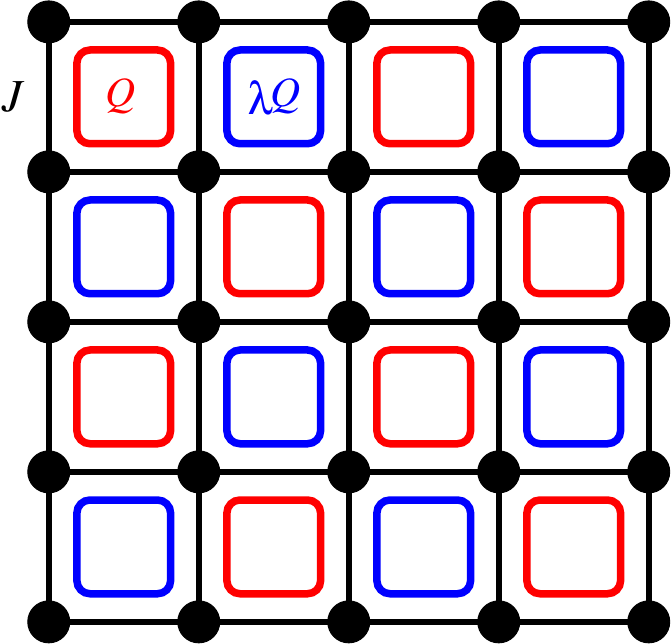}
  \caption{The extended CBJQ model. Black lines denote the Heisenberg interactions of strength $J$. The red $Q$-plaquettes, denoted as FP, represent the four-spin interactions of strength $Q$, and the blue $\lambda Q$-plaquettes, denoted as EP, represent the four-spin interactions of strength $\lambda Q$, with $\lambda\le1$.}
  \label{model}
  \end{figure}

The Hamiltonian of the extended CBJQ model is defined as
\be
H=-J\sum_{\la ij\ra}P_{ij}-Q\sum_{ijkl\in\square}(P_{ij}P_{kl}+P_{ik}P_{jl})\nonumber\\
-\lambda Q\sum_{ijkl\in\square^\prime}(P_{ij}P_{kl}+P_{ik}P_{jl})
\ee
where $P_{ij}=1/4-{\bm S}_i\cdot{\bm S}_j$ is the singlet projection operator on sites $i$ and $j$, $\la ij\ra $ denotes the nearest-neighbor bonds on a periodic square lattice with $N=L^2$ sites. The symbol $\square$ refers to the red $2\times2$ $Q$-plaquettes, and $\square^\prime$ denotes the blue $2\times2$ $\lambda Q$-plaquettes, as illustrated in Fig.~\ref{model}. We refer to these two types of plaquettes as the full plaquette (FP) and the empty plaquette (EP), respectively. The antiferromagnetic Heisenberg interaction $J$ is set to unity to fix the energy scale. The four-spin coupling strength is $Q$ on the FP and $\lambda Q$ on the EP. The parameter $\lambda$ controls the relative strength of the four-spin interactions on the two types of plaquettes. 
Since the parameter sets $(\lambda, Q)$ and $(1/\lambda, \lambda Q)$ are physically equivalent in the ground state, we restrict our study to the range $0 \le \lambda \le 1$.

When $\lambda=0$, the model reduces to the original CBJQ model, which exhibits a single first-order quantum phase transition at $Q_c=4.5977(1)$ from the PS state on the FP to the AFM state~\cite{Bzhao2019}. 
We refer to this PS state localized on the FP as the full plaquette-singlet (FPS) state, and correspondingly denote the PS state on the EP as the empty plaquette-singlet (EPS) state. In the CBJQ model, this transition is first order and is accompanied by an emergent $O(4)$ symmetry enhancement.

When $\lambda=1$, the model corresponds to the conventional two-dimensional $J-Q_2$ model, which undergoes a quantum phase transition at $Q_c=22.381$ from the AFM phase to the fourfold-degenerate columnar valence-bond solid (VBS) phase~\cite{jq2-sandvik,science-shao}. At the transition point, the model exhibits an emergent $O(5)$ symmetry~\cite{nahum2015,deng2024}. The transition has been further shown to be weakly first-order and located close to the $SO(5)$ quantum critical point~\cite{deng2024,jun2024}.

To characterize the ordered phases, we first introduce the order parameters. As the SSE method is more efficient for $S^z$-basis observables, the order parameters used in this paper are all defined solely in terms of the $S^z$ spin components.
The order parameters for FPS and AFM phases are defined as
\be
m_p=\frac{2}{N}\sum_{{\bm q}}m_p({\bm q}),\quad m_s=\frac{1}{N}\sum_i m_s(i),
\ee
where $m_p({\bm q})=\theta({\bm q})\Pi({\bm q})$ is the local PS order parameter on a FP. Here, $\theta({\bm q})=\pm1$ distinguishes even and odd plaquette rows, and $\Pi({\bm q})=S^z({\bm q}_1)S^z({\bm q}_2)S^z({\bm q}_3)S^z({\bm q}_4)$ with ${\bm q}_i (i=1,2,3,4)$ labeling the four corner sites of the FP. The local AFM order parameter on site $i$ is $m_s(i)=\phi(i)S^z_i$, where $\phi(i)$ denotes the staggered AFM phase factor.

 The VBS order parameter ${\bm D}=(D_x,D_y)$ can also be defined in terms of the $S^z$ spin components as
\be
D_\mu=\frac{1}{N}\sum_{i=1}^{N}\psi(i)S_i^zS_{i+\hat{e}_\mu}^z, \quad \mu\in\{x,y\},
\ee
where $i+\hat{e}_\mu$ denotes the nearest neighbor of site $i$ in the positive $\mu$ direction, $\psi(i)=(-1)^{\mu_i}$ is the VBS phase factor, and $\mu_i\in\{x_i,y_i\} $ represents the corresponding spatial coordinate of site $i$.

As $\lambda$ approaches $1$, the FPS and EPS states are expected to compete strongly, potentially leading to the emergence of EPS order. To characterize this phase, we introduce the EPS order parameter $m_e$ defined as
\be
m_e=\frac{2}{N}\sum_{{\bm q^\prime}}m_e({\bm q^\prime}),
\ee
where $m_e({\bm q^\prime})=\theta({\bm q^\prime})\Pi({\bm q^\prime})$ denotes the local PS order parameter on an EP. Here, $\theta({\bm q^\prime})=\pm1$ distinguishes even and odd plaquette rows, and $\Pi({\bm q^\prime})=S^z({\bm q^\prime}_1)S^z({\bm q^\prime}_2)S^z({\bm q^\prime}_3)S^z({\bm q^\prime}_4)$, with ${\bm q^\prime}_i (i=1,2,3,4)$ labeling the four corner sites of the EP.

The corresponding Binder ratios for the four order parameters are defined as
\be
U_p&=&\frac{\la m_p^4\ra}{\la m_p^2\ra^2},~~U_s=\frac{\la m_s^4\ra}{\la m_s^2\ra^2},\nonumber\\
U_{VBS}&=&\frac{\la {\bm D}^4\ra}{\la {\bm D}^2\ra^2},~~U_{ep}=\frac{\la m_e^4\ra}{\la m_e^2\ra^2},
\ee
respectively. The transition points can be located using these Binder ratios.

\section{Numerical results}
\label{results}
\subsection{Phase diagram}
\begin{figure}[t]
  \includegraphics[width=0.45\textwidth,clip]{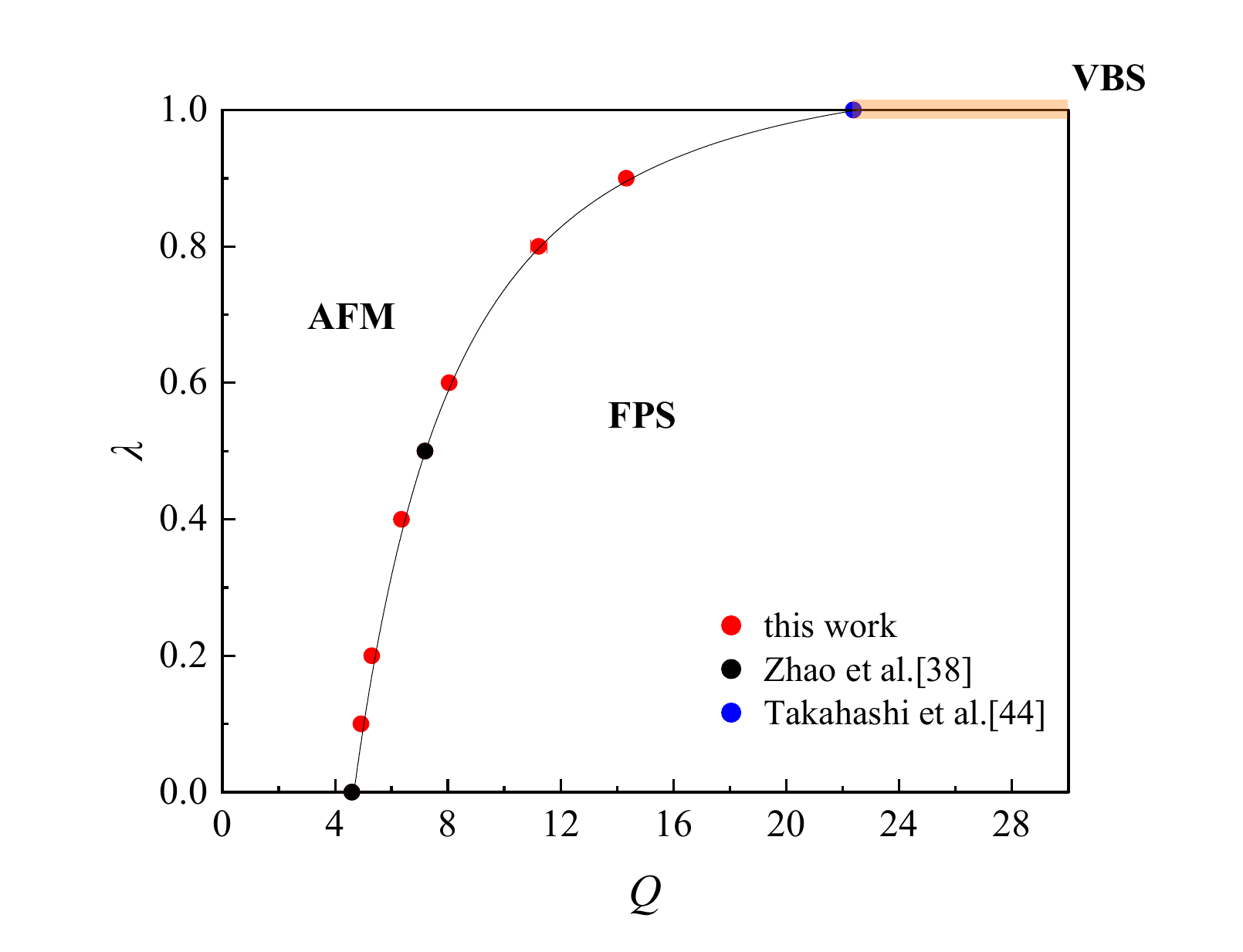}
  \caption{Phase diagram of the extended model in the $(Q,\lambda)$-plane. Transitions in this phase diagram are all first-order. The black and blue points are selected estimates from previous studies. The solid line is guide for eye.}
  \label{phase}
  \end{figure}

We first present the phase diagram of the extended CBJQ model in the $(Q,\lambda)$-plane.
 As shown in Fig.~\ref{phase}, for $\lambda<1$ the model possesses only two ground-state phases: the AFM phase and the FPS phase. A direct first-order phase transition separates these two phases, and the system exhibits an emergent $O(4)$ symmetry on the transition line. When $\lambda=1$, the model deduces to the $J-Q_2$ model~\cite{jq2-sandvik}, which hosts a weakly first-order phase transition from AFM to 
 VBS and exhibits emergent $O(5)$ symmetry at the transition point.
 
\begin{figure}[t]
  \includegraphics[width=0.5\textwidth,clip]{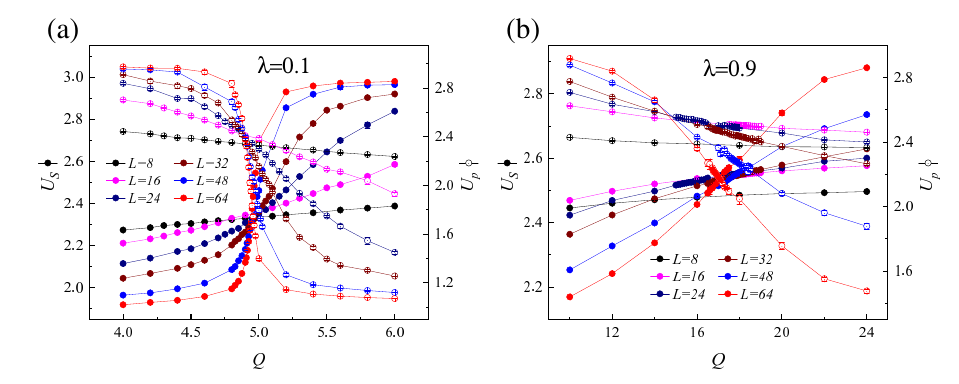}
  \caption{Binder ratios $U_s$ and $U_p$ of the extended CBJQ model versus $Q$ for different system sizes at (a) $\lambda=0.1$ and (b) $\lambda=0.9$.}
  \label{usup-1-9}
  \end{figure}

The transition points are estimated using the Binder ratios. If a phase transition exists, the corresponding Binder ratios for different system sizes will converge to a common crossing point at the transition as the system size increases. We study the crossing points of curves for two different system sizes, $L$ and $2L$, by locating the value of $Q$ at which $U(2L)$ and $U(L)$ intersect. From the analysis of these crossing points, we find that there is a single transition for all $\lambda<1$. In Fig. \ref{usup-1-9}, we show the Binder ratios $U_s$ and $U_p$ for $\lambda=0.1$ and $0.9$ as examples. In both cases, the two Binder ratios exhibit the same crossing point, which is similar to the $\lambda=0$ case~\cite{Bzhao2019} and indicates a single transition. A more detailed analysis of the crossing points for $\lambda=0.9$ is provided in Sec.~\ref{results}~C. The Binder ratios $U_s$ and $U_p$ for other $\lambda$ are shown in the Appendix.

\begin{figure}[t]
  \includegraphics[width=0.45\textwidth,clip]{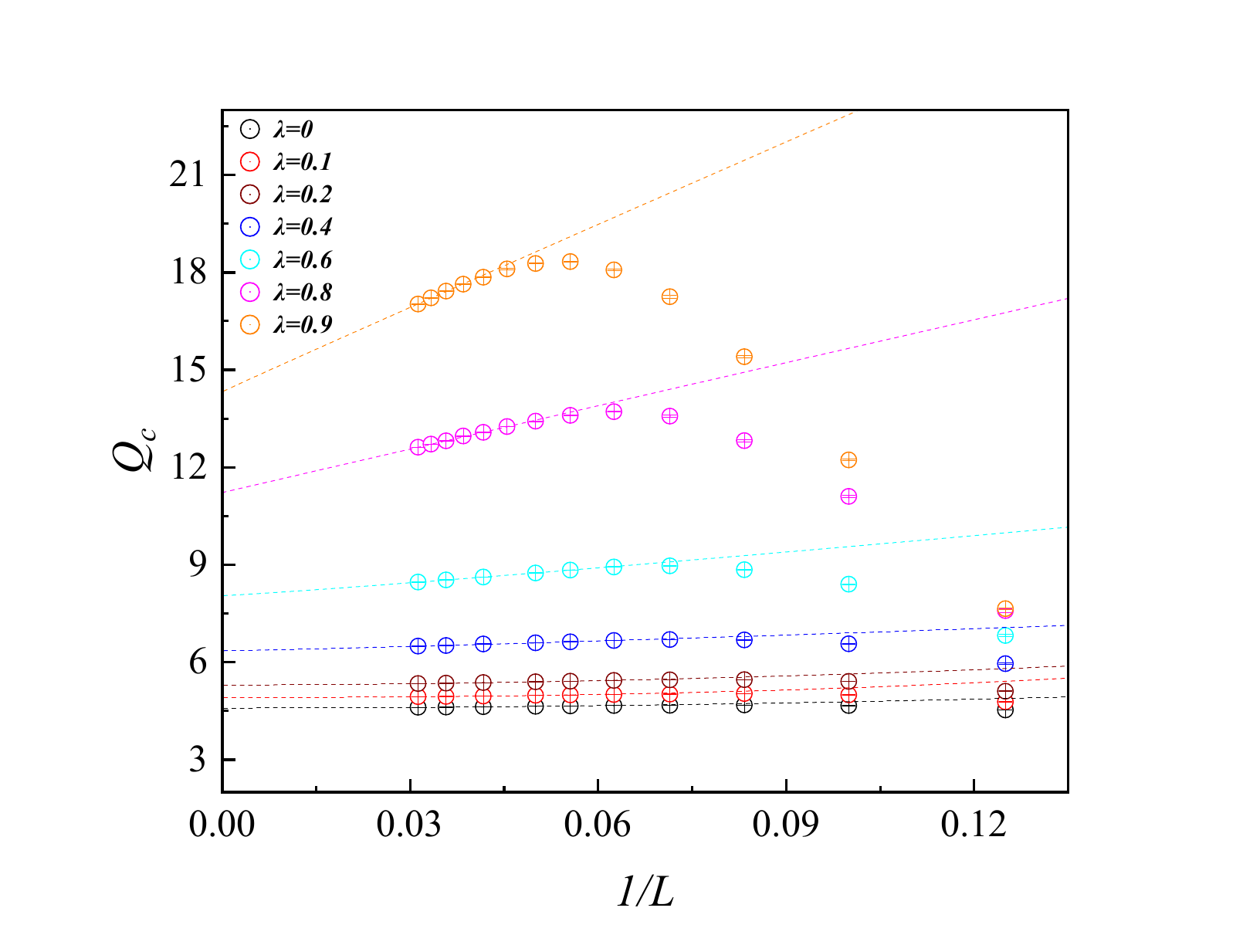}
  \caption{Critical points extracted from the crossing points of $U_s(L)$ and $U_s(2L)$. The dashed lines for $\lambda\le0.8$ are obtained from individual fits, while the dashed line at $\lambda=0.9$ is obtained from a joint fit.}
  \label{allqc}
  \end{figure}

In this paper, we focus primarily on the crossing points of $U_s$. Due to limited computational resources, we do not compute the crossing points of $U_p$ except at $\lambda=0.9$, where the nonmonotonic finite-size effect is pronounced. In Fig.~\ref{allqc}, we present the crossing point $Q_c(L)$ of $U_s(L)$ and $U_s(2L)$ as functions of $1/L$. For $\lambda\le 0.6$, we extract the transition points $Q_c$ via power-law fit to these crossing points. For $\lambda=0.8$, the nonmonotonic behavior is too severe to allow a reliable power-law fit; instead, we perform a polynomial fit and take the value extrapolated to $1/L=0$ as $Q_c$. For $\lambda=0.9$, the nonmonotonic behavior becomes even more pronounced, rendering any simple fit unreliable. We therefore determine $Q_c$ using a joint fitting procedure, which is discussed in Sec.~\ref{results}~C.

  \begin{figure}[t]
  \includegraphics[width=0.5\textwidth,clip]{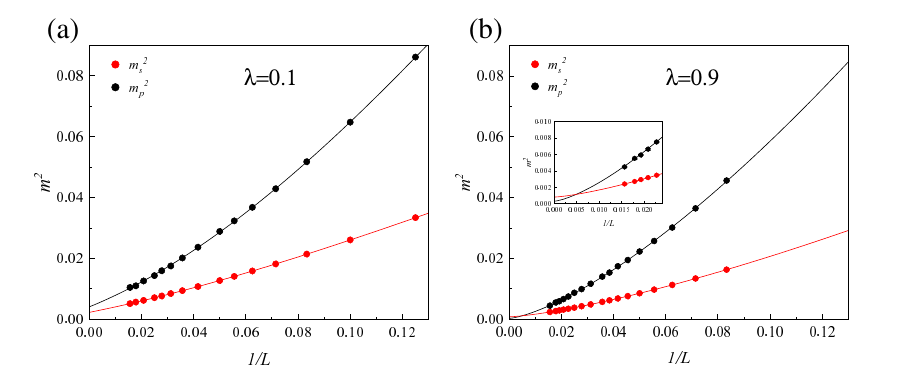}
  \caption{Squared order parameters $m_s^2$ and $m_p^2$ plotted versus $1/L$ at the transition point $Q_c$ for (a) $
  \lambda=0.1$ and (b) $\lambda=0.9$. The inset in (b) shows a zoom of the large‑system‑size region. Solid lines represent the fit results.}
  \label{msmp}
  \end{figure}
  \begin{figure}[t]
  \includegraphics[width=0.5\textwidth,clip]{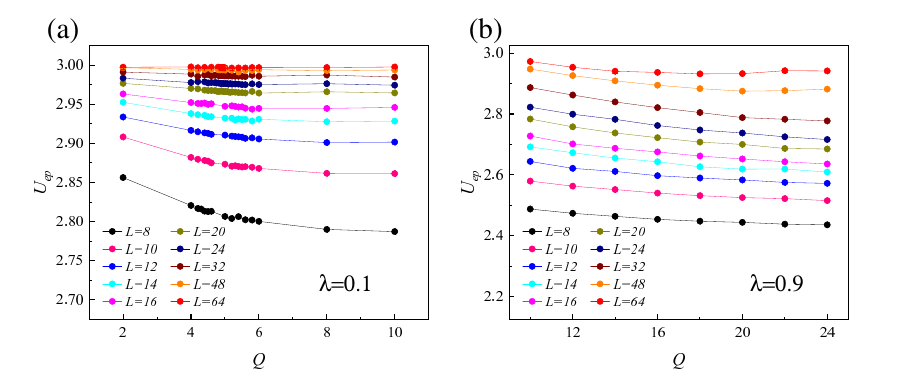}
  \caption{Binder ratio $U_{ep}$ versus $Q$ for different system sizes at (a) $\lambda=0.1$ and (b) $\lambda=0.9$.}
  \label{uep}
  \end{figure}
Figure~\ref{msmp} presents the squared order parameters 
$m_s^2$ and $m_p^2$ at the transition point $Q_c$ for $\lambda=0.1$ and $0.9$ as examples. For $\lambda=0.1$, we employ polynomial fits, which extrapolate to finite values for both order parameters in the thermodynamic. For $\lambda=0.9$, we expect the AFM-FPS transition is weakly first-order, and the order parameters are very small at the transition point. The transition point is close to an $SO(5)$ multicritical point, akin to that of the $J-Q_2$ model~\cite{jun2024}. Consequently, critical fluctuations dominate over a wide range of length scales. Therefore, we use power-law fits to extract the thermodynamic limit of the order parameter, as the finite-size data exhibit approximate critical scaling before eventually crossing over to the asymptotic first-order behavior. In both cases, the extrapolated values of the two order parameters remain finite in the thermodynamic limit, confirming the coexistence of AFM and FPS order. Thus, even though power-law fits are used for $\lambda=0.9$ due to its proximity to the SO(5) multicritical point, the transition is first-order. The same coexistence behavior is observed for all other $\lambda$ parameters studied, as shown in the Appendix. We therefore conclude that the transition from the AFM to FPS phase is first-order for all $\lambda$. 

To examine the possible existence of an EPS phase in the phase diagram, we calculate the Binder ratio $U_{ep}$ for the EPS order parameter. Figure~\ref{uep} shows $U_{ep}$ for $\lambda=0.1$ and $0.9$ as examples. For both parameters, no crossing-point behavior is observed, indicating the absence of an EPS phase. Additional results for other $\lambda$ values (presented in the Appendix) exhibit the same behavior. Based on these observations, we conclude that there is no EPS phase in the phase diagram.

\subsection{Emergent symmetry at the transition points}
\begin{figure*}[htbp]
  \includegraphics[width=0.95\textwidth,clip]{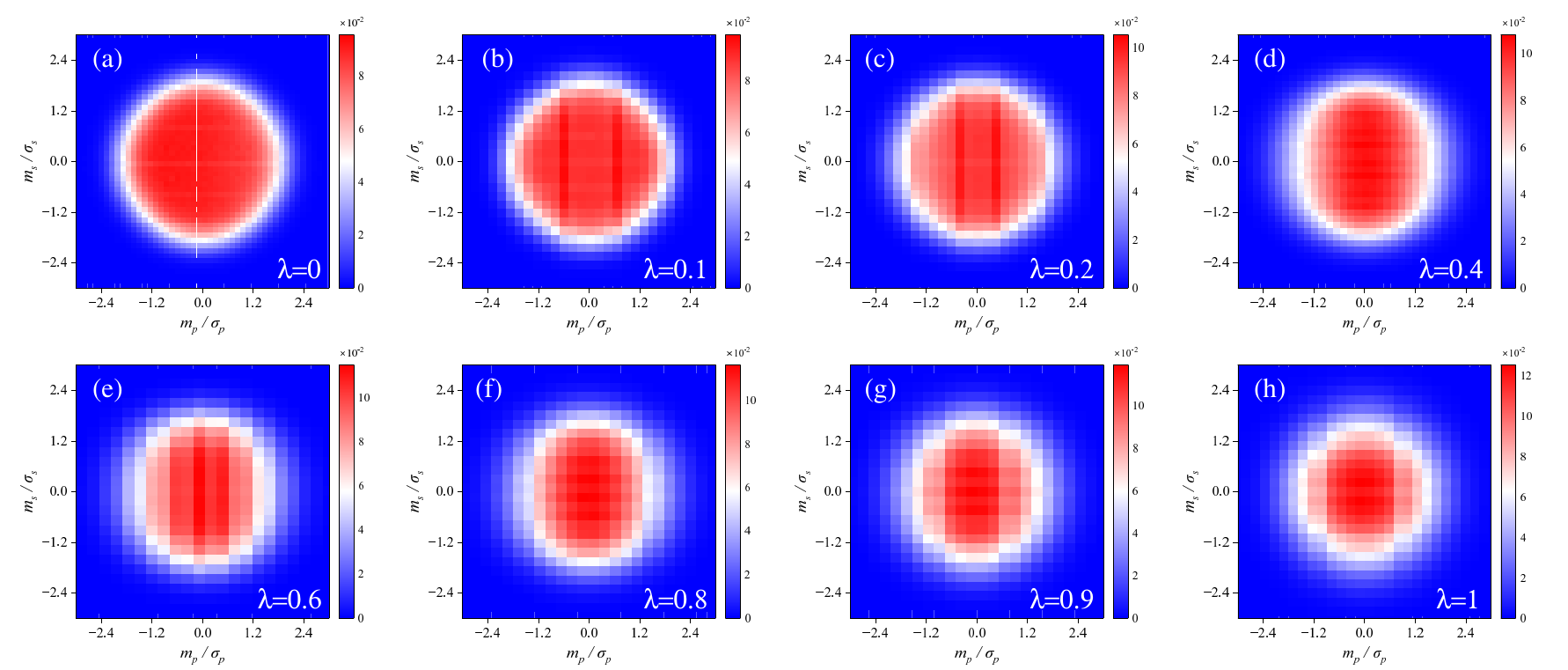}
  \caption{Distribution $P(m_p/\sigma_p,m_s/\sigma_s)$ at the transition point for different $\lambda$. The system size is $L=64$.}
  \label{mpmzhist}

  \end{figure*}
    \begin{figure*}[htbp]
  \includegraphics[width=0.95\textwidth,clip]{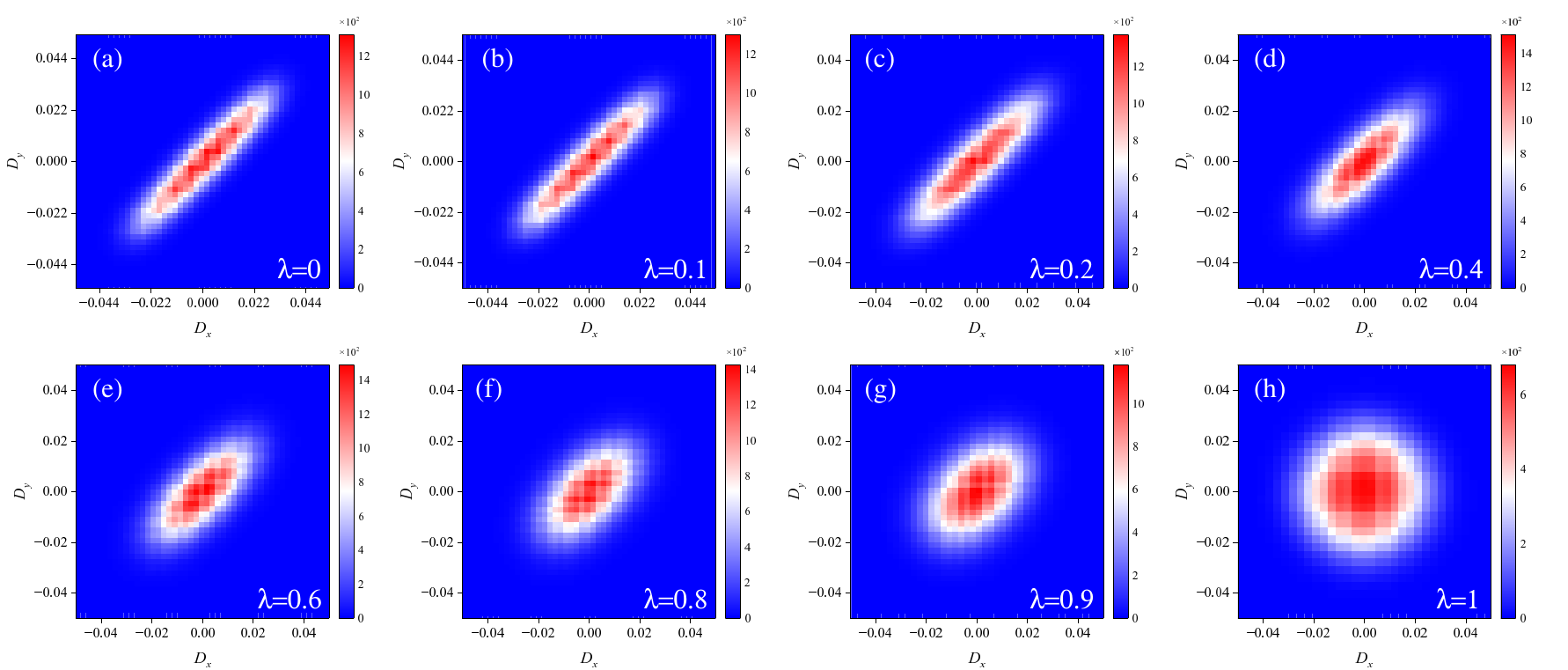}
  \caption{Distribution $P(D_x,D_y)$ at the transition point for different $\lambda$. The system size is $L=64$.}
  \label{dxdyhist}
  \end{figure*}
The CBJQ model ($\lambda=0$) exhibits an emergent $O(4)$ symmetry at the transition~\cite{Bzhao2019}, while the $J-Q_2$ model ($\lambda=1$) features an emergent $O(5)$ symmetry~\cite{deng2024}. Here we investigate whether the $O(4)$ symmetry persists along the entire transition line for $\lambda<1$, or whether a multicritical point exists where the symmetry changes from $O(4)$ to $O(5)$.

To test the symmetry at the transition points, we accumulate the distribution $P(m_p,m_s)$. 
For points uniformly distributed on an $O(4)$ sphere, the projection onto the two components should yield a uniform distribution inside a circle~\cite{Bzhao2019}. In contrast, for points uniformly distributed on an $O(5)$ sphere, the projection onto two components gives a different distribution, where the probability density peaks at zero. The $O(5)$ case is illustrated in the Appendix. To make the putative $O(4)$ symmetry manifest, we normalize $m_p$ and $m_s$ by the factors $\sigma_p=\sqrt{\la m_p^2\ra}$ and $\sigma_s=\sqrt{\la m_s^2\ra}$.

Distributions $P(m_p/\sigma_p,m_s/\sigma_s)$ for all simulated $\lambda$ are shown in Fig.~\ref{mpmzhist}. For $\lambda\le0.8$, a uniform distribution inside a circle is clearly observed. For $\lambda=1$, the distribution resembles that of $O(5)$ case. For $\lambda=0.9$, it is not possible to determine whether the symmetry is $O(4)$ or $O(5)$.  
To resolve this ambiguity, we additionally accumulate the distribution $P(D_x,D_y)$ at the transition points, as shown in Fig.~\ref{dxdyhist}. The twofold-degenerate FPS phase produces two peaks in $P(D_x,D_y)$ along the $(1,1)$ drection, while the EPS phase has two peaks along the $(1,-1)$ direction. We refer to the $(1,1)$ direction as the "FPS" direction and the $(1,-1)$ direction as the "EPS" direction. At the transition, unlike the rotationally invariant case in $P(m_p/\sigma_p,m_s/\sigma_s)$, the distribution exhibits directional preferences for $\lambda<1$: it takes large values along the FPS direction and small values along the EPS direction.
As we have found that $m_p$ remains finite in the thermodynamic limits while $m_e$ vanishes, we expect that the distribution will collapse onto a line along the FPS direction for all $\lambda<1$.
For $\lambda=1$, however, the distribution becomes $U(1)$-rotationally invariant with respect to $D_x$ and $D_y$, which are precisely two components of the $O(5)$ symmetry. We therefore conclude that for $\lambda<1$ the system retains $O(4)$ symmetry, and only at $\lambda=1$ does the $O(5)$ symmetry emerge.

For the CBJQ model, the emergent $O(4)$ symmetry at the transition arises from the combination of the $O(3)$ symmetry of the AFM order and the $Z_2$ symmetry of the FPS order. In the transition region, the system can be describe by an effective deformed quantum $O(4)$ model (the XXXZ model), and the control parameter $Q$ tunes the order parameter from the $O(3)$ phase through the $O(4)$ point into the $Z_2$ phase~\cite{Bzhao2019}. For the $J-Q_2$ model, the emergent $O(5)$ symmetry at the transition arises from the combination of the $O(3)$ symmetry of the AFM order and the $Z_4$ symmetry of the VBS order. For the extended CBJQ model, we expect that $(m_p,m_{ep})$ plays the same role as  $(D_x,D_y)$ in the $O(5)$ framework. Consequently, along the transition line the system can be described by an effective deformed quantum $O(5)$ model (the XXXXZ model) and the control parameter $(\lambda,Q_c)$ tunes the order parameter from the $O(4)$ phase into the $O(5)$ phase.

\begin{figure}[t]
  \includegraphics[width=50mm,clip]{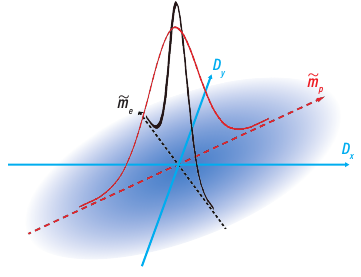}
  \caption{Schematic distribution $P(D_x,D_y)$ for the extended CBJQ model in the AFM phase. The red (black) dashed axis represents the FPS (EPS) direction $\widetilde{m}_p(\widetilde{m}_e)$, and the distribution $P(D_x,D_y)$ is a product of two independent Gaussian distributions of $\widetilde{m}_p$ and $\widetilde{m}_e$.}
  \label{gauss}
  \end{figure}

\begin{figure}[t]
  \includegraphics[width=0.5\textwidth,clip]{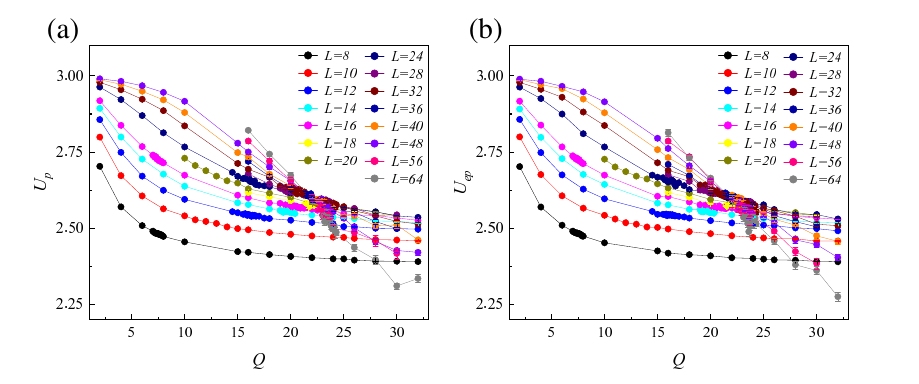}
  \caption{Binder ratios for the $J-Q_2$ model: (a) $U_{p}$ and (b) $U_{ep}$.}
  \label{ujq2}
  \end{figure}

Such emergent symmetry from $O(4)$ to $O(5)$ can also be observed from the Binder ratios $U_p$ and $U_{ep}$. As we have defined two directions in the $(D_x,D_y)$ plane (the FPS direction and the EPS direction), we use these two directions as new axes, denoted $\widetilde{m}_p$ and $\widetilde{m}_{e}$, as shown in Fig.~\ref{gauss}. These serve as the effective order parameters corresponding to $m_p$ and $m_e$, respectively. As shown in Fig.~\ref{ujq2}, for $\lambda=1$, $U_p$ and $U_{ep}$ exhibit identical behavior, which is significantly different from that for $\lambda<1$. This indicates an emergent symmetry between the FPS order and the EPS order at $\lambda=1$. The emergent $O(5)$ symmetry from $O(4)$ thus originates from this new symmetry between the FPS and EPS order.

\subsection{Effective order parameter}

\begin{figure}[t]
  \includegraphics[width=0.45\textwidth,clip]{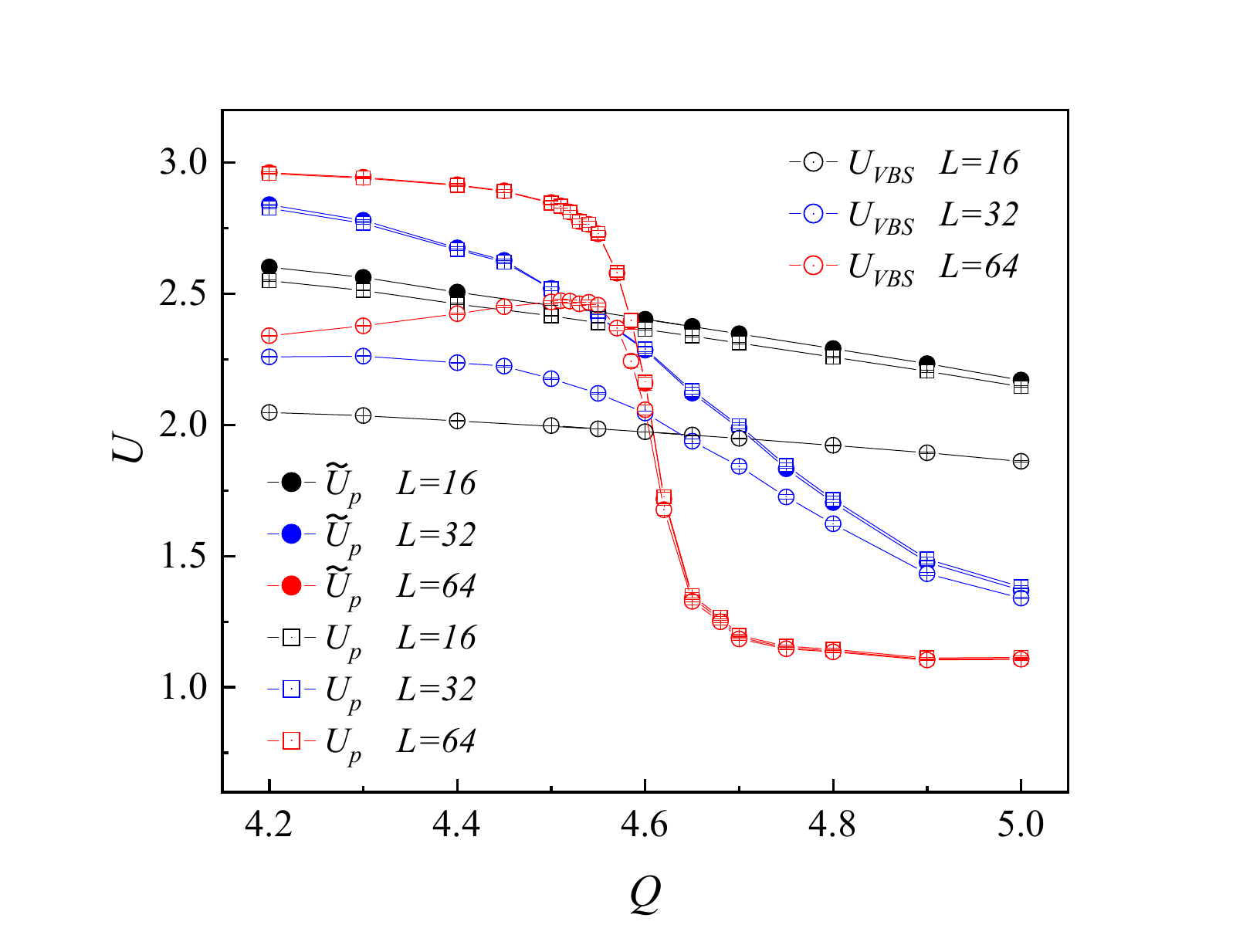}
  \caption{Binder ratios $U_p$, $U_{VBS}$, and $\widetilde{U}_p$ in the CBJQ model ($\lambda=0$).}
  \label{binders}
  \end{figure}

The phase transition from the AFM to FPS phase in the extended CBJQ model can also be characterized by the VBS order parameter. As mentioned above, we define two directions in the $(D_x,D_y)$ plane: the FPS direction $\widetilde{m}_p$ and the EPS direction $\widetilde{m}_e$.
In the FPS phase, the distribution $P(D_x,D_y)$ exhibits two peaks at nonzero values along the FPS direction. In the AFM phase, $P(D_x,D_y)$ is concentrated at zero~\cite{Bzhao2019}. At the transition, $P(D_x,D_y)$ takes nonzero values along the FPS direction within a finite region centered around zero. This behavior can be used to characterize the phase transition to the FPS phase. The Binder ratio $U_{VBS}$ also exhibits singular behavior at the transition point. As an example, $U_{VBS}$ for $\lambda=0$ is shown in Fig.~\ref{binders}. The singular behavior occurs at the same point as that of the FPS Binder ratio $U_p$ in the thermodynamic limit. Thus, $U_{VBS}$ can be treated as an effective order parameter for the FPS phase. Binder ratios $U_{VBS}$ for all $\lambda$ values are presented in the Appendix.

\begin{figure}[t]
  \includegraphics[width=0.45\textwidth,clip]{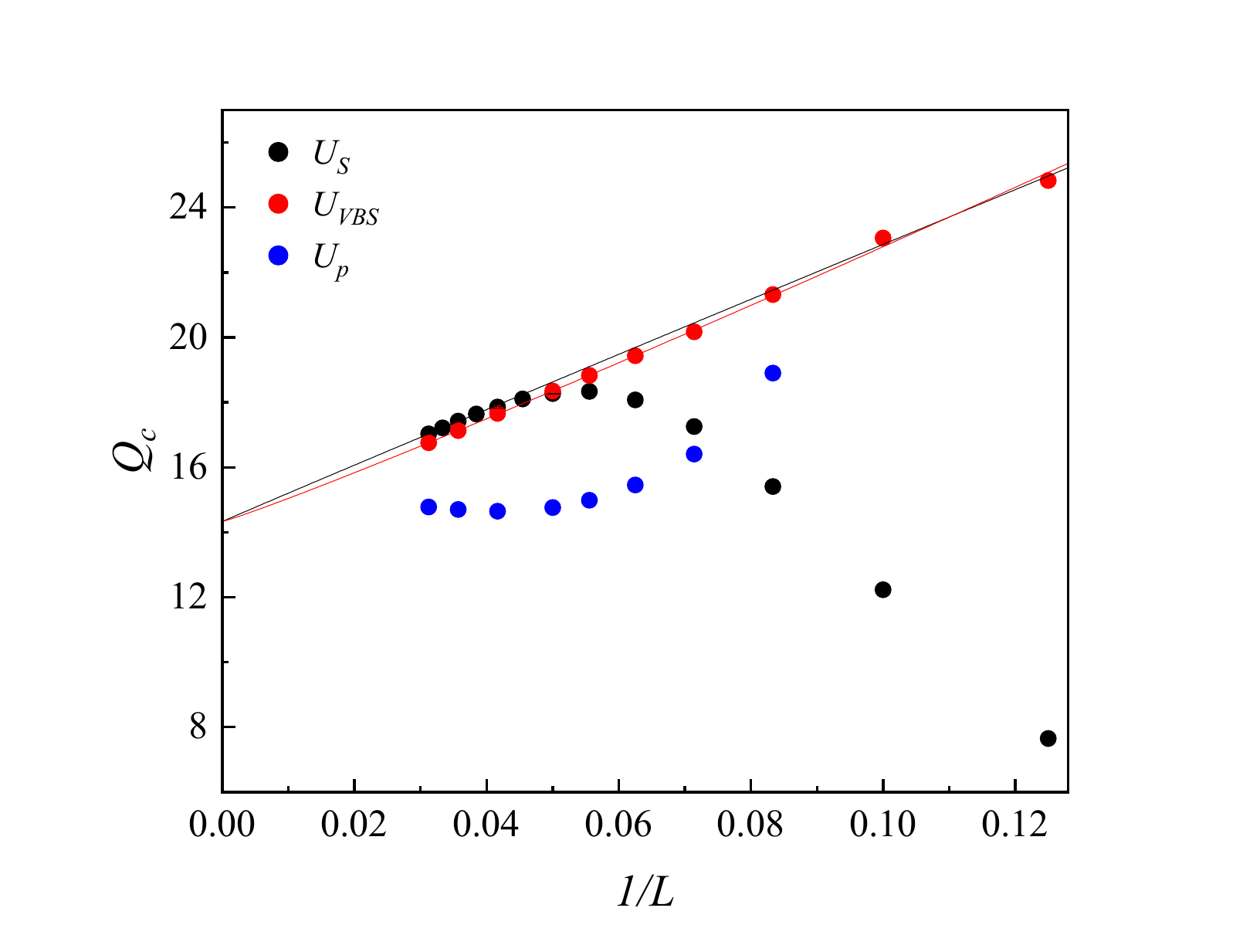}
  \caption{Crossing points of $U_s$, $U_{VBS}$, and $U_p$ at $\lambda=0.9$. The solid lines represent the joint-fit results for $U_s$ and $U_{VBS}$.}
  \label{jointfit}
  \end{figure}

We study the crossing points of $U_{VBS}$ for system sizes $L$ and $2L$ to locate the transition point of the FPS phase at $\lambda=0.9$ as an example, where the finite-size effect is particularly severe. We present the crossing points of $U_s$, $U_{VBS}$ and $U_p$ at $\lambda=0.9$ in Fig.~\ref{jointfit}. The crossing points of $U_s$ and $U_p$ exhibit a nonmonotonic behavior even for very large systems, making a reliable fit difficult. In contrast, the crossing points of $U_{VBS}$ are monotonic over the entire range of system sizes studied. Transition point extracted from $U_{VBS}$ is more reliable. As shown in the figure, the crossing point behavior of the three Binder ratios indicates a single transition point. We therefore perform a joint fit of $U_s$ and $U_{VBS}$, which replace $U_p$ with $U_{VBS}$, to extract the transition point.


In addition to its role in locating the transition, we observe an intriguing feature of $U_{VBS}$ for $\lambda\le0.8$. In this parameter regime, the Binder ratio $U_{VBS}$ on larger lattices exhibits a gradual increase with $Q$ within the AFM phase. Results for $\lambda = 0$ are shown in Fig. \ref{binders}; the corresponding data for other $\lambda$ values are presented in the Appendix. This gradual increase might be interpreted as a weak first-order transition, where $U_{VBS}$ is expected to diverge at the transition point in the thermodynamic limit. However, we argue that this behavior is not a signature of a first-order transition but rather arises from an additional irrelevant field.

Figure~\ref{gauss} presents a schematic distribution $P(D_x,D_y)$ for the extended CBJQ model in the AFM phase. The $(D_x,D_y)$ plane can be decomposed into a combination of the $\widetilde{m}_p$ and $\widetilde{m}_e$ directions. The $\widetilde{m}_p$ direction corresponds to the FPS order, and the $\widetilde{m}_e$ direction corresponds to the EPS order. In the AFM phase, both the FPS order and the EPS order are absent. The distribution $P(D_x,D_y)$ can be regarded as the product of two independent Gaussian distributions along $\widetilde{m}_p$ and $\widetilde{m}_e$ directions. 
The effective order parameter $\widetilde{m}_p$ (for FPS) and $\widetilde{m}_e$ (for EPS) can be constructed by projecting $(D_x, D_y)$ onto the respective axes:
\begin{equation}
\widetilde{m}_p = \frac{D_x + D_y}{\sqrt{2}}, \quad \widetilde{m}_e = \frac{D_y - D_x}{\sqrt{2}}.
\end{equation}
Binder ratios $\widetilde{U}_p$ for the effective order parameter $\widetilde{m}_p$ are presented in Fig. \ref{binders}. For $L=64$, $U_p$ and $\widetilde{U}_p$ are identical, indicating that $\widetilde{m}_p$ indeed behaves as the effective order parameter for the FPS phase.

The variances of these two Gaussian distributions are generally unequal, and the Binder ratio $U_{VBS}$ depends on the ratio of these variances: $x=\sigma^2(\widetilde{m}_p)/\sigma^2(\widetilde{m}_e)$.
In the AFM phase, the $U_{VBS}$ can be expressed as $(3x^2 + 2x + 3)/(x+1)^2$. At $Q = 0$, the two variances are equal, yielding $U_{VBS} = 2$. As $Q$ increases, the variance ratio grows, and $U_{VBS}$ increases accordingly. 
In the limit $Q \to Q_C$, the variances ratio diverges in the thermodynamic limit, and we obtain $U_{VBS} = U_p \to 3$. Consequently, in the thermodynamic limit, $U_{VBS}$ increases from $2$ to $3$ within the AFM phase. For sufficiently large but finite system sizes, $U_{VBS}$ also increases with $Q$ but cannot reach the limiting value of $3$. As shown in Fig.~\ref{binders}, for $L=64$, $U_{VBS}$ increases when $Q<Q_c$. For all other $\lambda$ values except $\lambda = 0.9$, $U_{VBS}$ exhibits similar behavior, as shown in the Appendix. The absence of this behavior at $\lambda = 0.9$ is likely due to a crossover from $O(5)$ to $O(4)$ symmetry. We expect that, for sufficiently large system sizes, such behavior will eventually emerge for $\lambda=0.9$ as well.

\section{Summary}
\label{summary}
In this paper, we investigate the phase diagram of the extended CBJQ model, which interpolates between the CBJQ model ($\lambda=0$) and the conventional $J-Q_2$ model ($\lambda=1)$. Using large-scale QMC simulations, we map out the ground-state phase diagram in the $(Q,\lambda)$ plane. For $\lambda<1$, the system exhibits a direct phase transition between the AFM phase and the FPS phase. No evidence of an EPS phase is found for any $\lambda<1$.

We investigate the emergent symmetry at the transition. For $\lambda\le0.8$, the distribution of the normalized AFM and FPS order parameters is uniform inside a circle, signaling an emergent $O(4)$ symmetry. For $\lambda=1$, the distribution corresponds to the $O(5)$ symmetry. At $\lambda=0.9$, the behavior is ambiguous from the distribution alone. By analyzing the distribution $P(D_x,D_y)$, we resolve this ambiguity: for $\lambda<1$ the distribution shows a clear preference along the FPS direction, whereas at $\lambda=1$ it becomes $U(1)$-rotationally invariant. We conclude that the $O(4)$ symmetry persists for all $\lambda<1$, and the $O(5)$ symmetry emerges only at $\lambda=1$. There is no multicritical point in between.

Furthermore, we demonstrate that the VBS Binder ratio can serve as an effective order parameter for the FPS phase. The crossing points for VBS Binder ratio exhibit a monotonic behavior, enabling a reliable determination of the transition point. Within the AFM phase, the 
VBS Binder ratio increases gradually as the transition is approached. This increase, which might naively suggest a weak first-order transition, is explained by the additional irrelevant filed $\widetilde{m}_e$ and does not indicate a thermodynamic divergence.

Our work provides a comprehensive phase diagram of the extended CBJQ model and clarifies the nature of the symmetry enhancement in this family of models.

\begin{acknowledgments}
This work is supported by the National Natural Science Foundation of China under Grants No. 12304171 and Beijing Institute of Technology Research Fund Program for Young Scholars.
\end{acknowledgments}

\appendix
\section*{Appendix}
  \begin{figure}[t]
  \includegraphics[width=0.5\textwidth,clip]{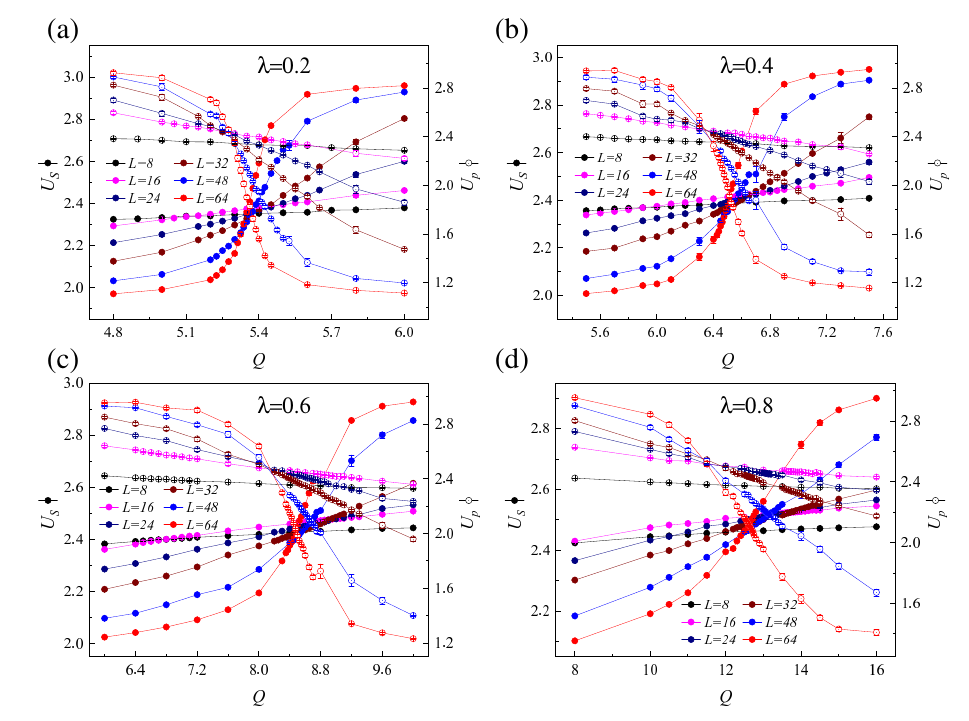}
  \caption{Binder ratios $U_s$ and $U_p$ for (a) $\lambda=0.2$, (b) $\lambda=0.4$, (c) $\lambda=0.6$, and (d) $\lambda=0.8$.}
  \label{usup-all}
  \end{figure}
Figure~\ref{usup-all} shows the AFM and FPS Binder ratios for $\lambda = 0.2, 0.4, 0.6$, and $0.8$. These two types of Binder ratios exhibit behavior similar to that of the CBJQ model, indicating a direct transition from AFM to FPS.

\begin{figure}[t]
  \includegraphics[width=0.5\textwidth,clip]{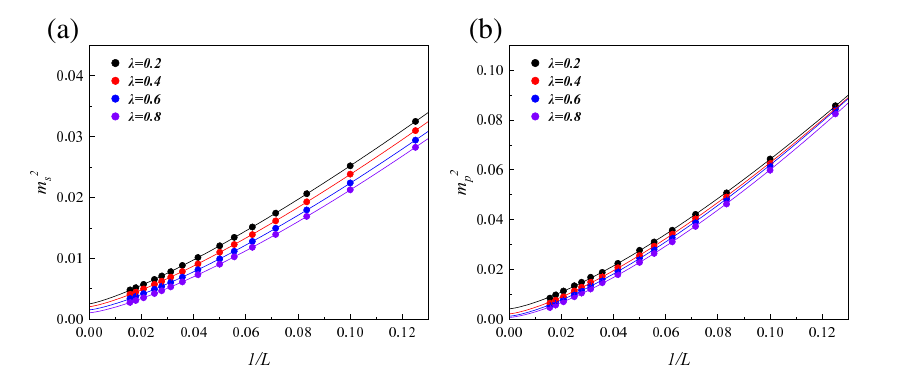}
  \caption{(a) The AFM order parameter $m_s^2$ and (b) the FPS order parameter $m_p^2$ for $\lambda=0.2$, $0.4$, $0.6$, and $0.8$. Solid lines represent the fit results.}
  \label{msmp2468}
  \end{figure}

Figure~\ref{msmp2468} shows the AFM and FPS order parameters for $\lambda=0.2$, $0.4$, $0.6$, and $0.8$. A power-law fit is used for all $\lambda$ values. All order parameters remain finite in the thermodynamic limit. Polynomial fit for small $\lambda$ yield similar results.

\begin{figure}[t]
  \includegraphics[width=0.5\textwidth,clip]{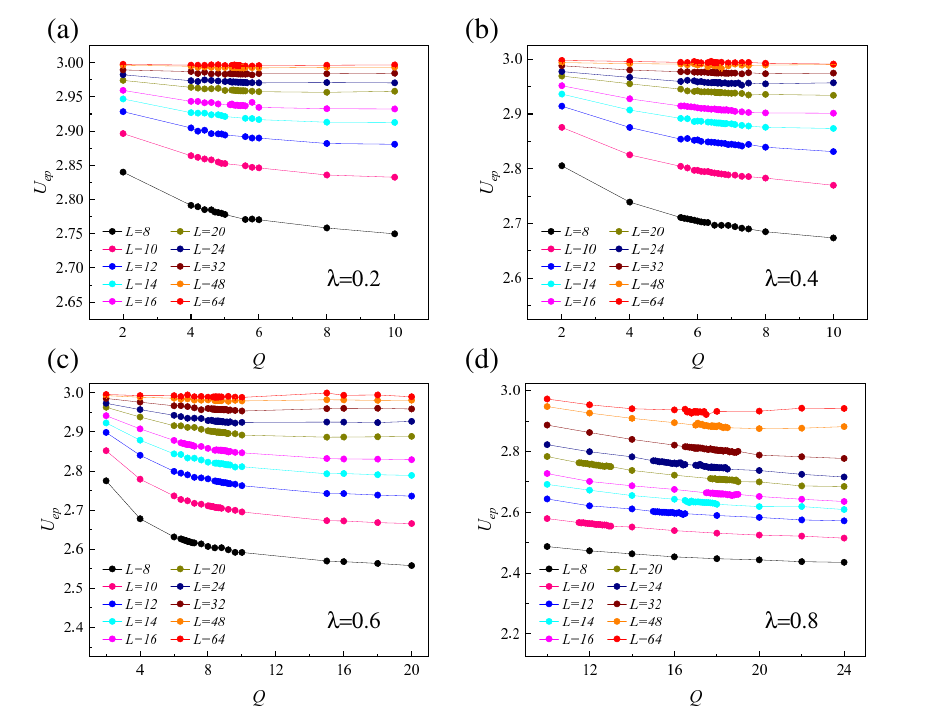}
  \caption{Binder ratios $U_{ep}$ for (a) $\lambda=0.2$, (b) $\lambda=0.4$, (c) $\lambda=0.6$, and (d) $\lambda=0.8$.}
  \label{uep2}
  \end{figure}

Figure~\ref{uep2} shows the EPS Binder ratios for $\lambda = 0.2, 0.4, 0.6$, and $0.8$. For each $\lambda$, the curves for different system sizes do not intersect, indicating the absence of an EPS phase in the phase diagram.

    \begin{figure}[t]
  \includegraphics[width=0.45\textwidth,clip]{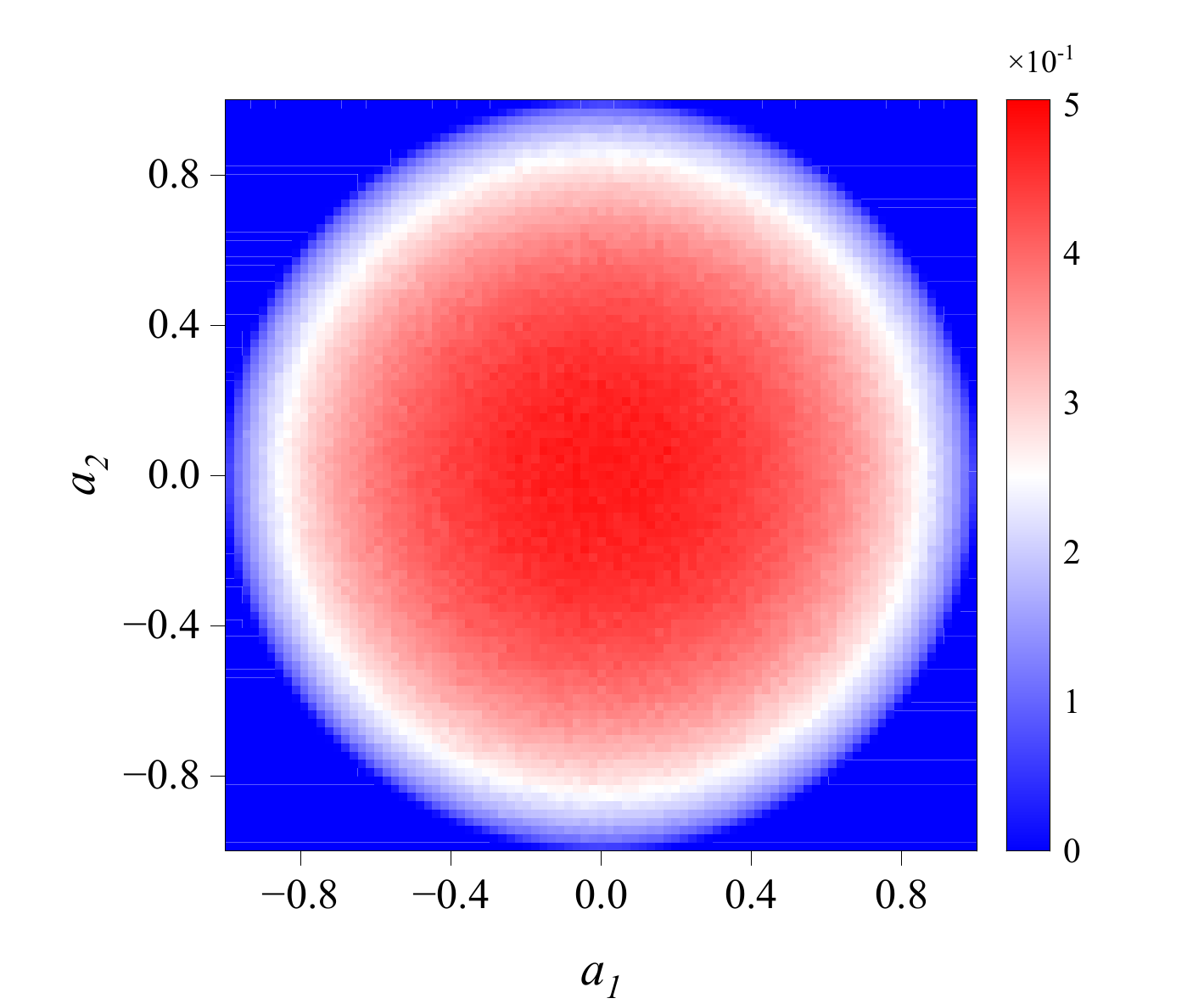}
  \caption{Two-component distribution for the $SO(5)$ symmetry.}
  \label{so5}
  \end{figure}

    \begin{figure}[t]
  \includegraphics[width=0.5\textwidth,clip]{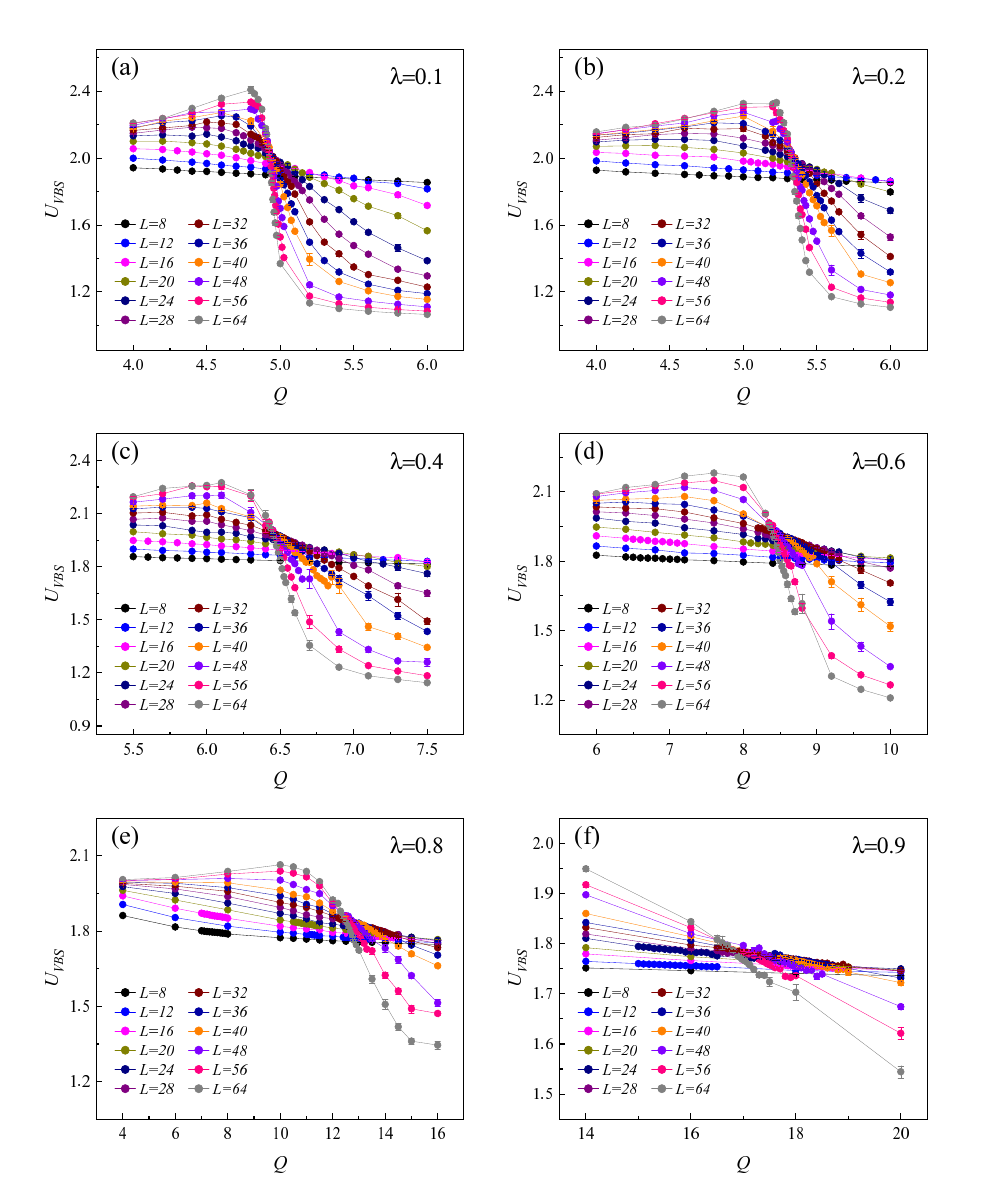}
      \caption{Binder ratios $U_{VBS}$ for (a) $\lambda=0.1$, (b) $\lambda=0.2$, (c) $\lambda=0.4$, (d) $\lambda=0.6$, (e) $\lambda=0.8$, and (f) $\lambda=0.9$.}
      \label{uvbs}
  \end{figure}

Figure~\ref{so5} shows the distribution obtained by projecting uniformly distributed points on an $O(5)$ sphere onto two of its components.

Figure~\ref{uvbs} presents the VBS Binder ratios $U_{VBS}$ for $\lambda = 0.1, 0.2, 0.4, 0.6, 0.8$, and $0.9$. For $\lambda \le 0.8$, $U_{VBS}$ gradually increases within the AFM phase. For $\lambda = 0.9$, however, no such increase is observed. We attribute this absence to a crossover behavior; we expect that such behavior will eventually appear when the system size becomes sufficiently large.

\end{document}